# Research on fuzzy PID Shared control method of small brain-controlled uav

Na Dong   Wen-qi Zhang and Zhong-ke Gao*
School of Electrical and Information Engineering, Tianjin University, Tianjin, China

**ABSTRACT:** Brain-controlled unmanned aerial vehicle (uav) is a uav that can analyze human brain electrical signals through BCI to obtain flight commands. The research of brain-controlled uav can promote the integration of brain-computer and has a broad application prospect.

At present, BCI still has some problems, such as limited recognition accuracy, limited recognition time and small number of recognition commands in the acquisition of control commands by analyzing eeg signals. Therefore, the control performance of the quadrotor which is controlled only by brain is not ideal. Based on the concept of Shared control, this paper designs an assistant controller using fuzzy PID control, and realizes the cooperative control between automatic control and brain control. By evaluating the current flight status and setting the switching rate, the switching mechanism of automatic control and brain control can be decided to improve the system control performance. Finally, a rectangular trajectory tracking control experiment of the same height is designed for small quadrotor to verify the algorithm.

## 1 introduction

Establishing a direct communication channel between the human brain and the outside world is an important concept [1]. Brain computer interface (BCI) is an information communication channel directly established between the brain and the computer or peripheral devices without relying on the conventional output pathway of the brain[2]. After sampling and analyzing human brain electrical signals, the idea of human control can be obtained, and then the idea of control can be exported to computers or peripherals, and finally the control can be realized. Although the research on BCI is still in its infancy, it has been proved to be effective in a wide range of applications, thus gaining significant development momentum. At present, the research of BCI mainly aims to help disabled people recover their mobility and communication ability, such as virtual keyboard [3]-[5], brain-controlled wheelchair [6][7], prosthesis [8] and so on. In addition, BCI also provides a revolutionary framework for designing real-time, brain-controlled systems [1]. Thus, there are researches on brain-controlled robots [9]-[12] and brain-controlled vehicles [13]-[15]. In recent years, there has been an upsurge of research on unmanned aerial vehicles (uavs) around the world. Uavs have a wide application prospect in both military and civil fields. At the same time, existing unmanned aerial vehicle (uav) systems are often unable to complete the mission successfully without human control and decision-making. In this context, the idea of developing a brain-controlled drone has led to a series of ongoing studies which are pioneered by researchers at the university of Minnesota [16]-[18].

At present, BCI still has some problems, such as limited recognition accuracy, limited recognition time and small number of recognition commands in the acquisition of control commands by analyzing eeg signals. Therefore, the control performance of the quadrotor which is controlled only by brain is not ideal. In this paper, based on the concept of Shared control, the fuzzy PID control is used to design the auxiliary controller to realize the cooperative control of automatic control and brain control. By evaluating the current flight status and setting the switching rate, the switching mechanism of automatic control and brain control is decided to improve the system control performance.

This paper is organized as follows:

1) system architecture is described in section 2

2) section 3 shows the result of attitude change of quadrotor controlled by brain control signals

3) section 4 USES fuzzy PID control to design auxiliary controller

4) at last, a rectangular trajectory tracking control experiment of the same height is designed for algorithm verification

## 2 brain-controlled uav system structure

In the brain-controlled uav system, people make decisions according to the aircraft state and environmental information and output control commands by BCI. The auxiliary fuzzy PID controller also outputs control instructions. By evaluating the current flight status and setting the switching rate, the switching mechanism of automatic control and brain control is decided. The system block diagram is as follows:

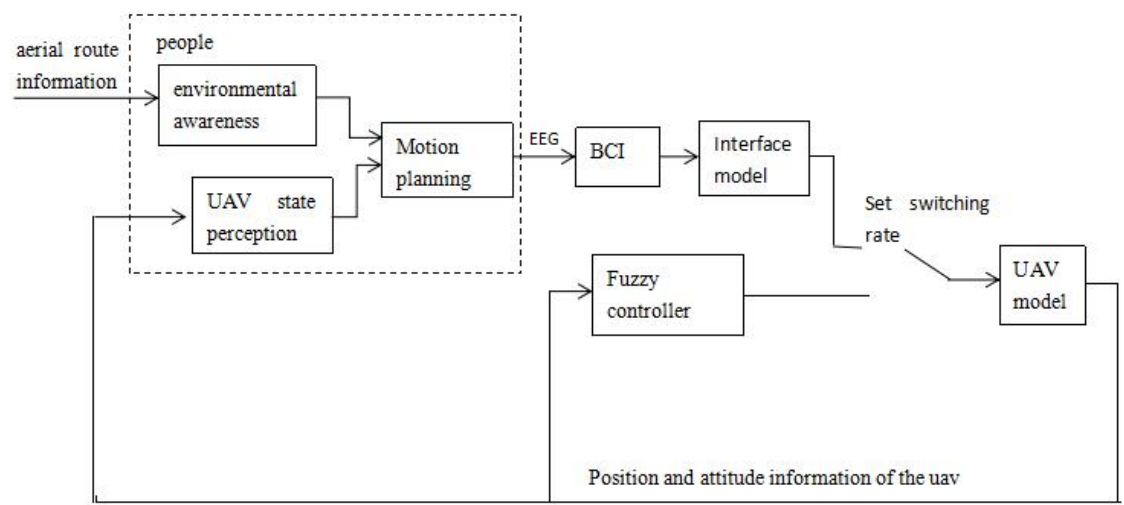

Fig 1 brain-controlled uav system structure

## 3 brain control and uav attitude control fusion

With the electric quadrotor taken as the controlled object, the control signals that are extracted by brain are used to control the motor speed change of each rotor, and the attitude control of the uav can be realized. It is shown in the figure.

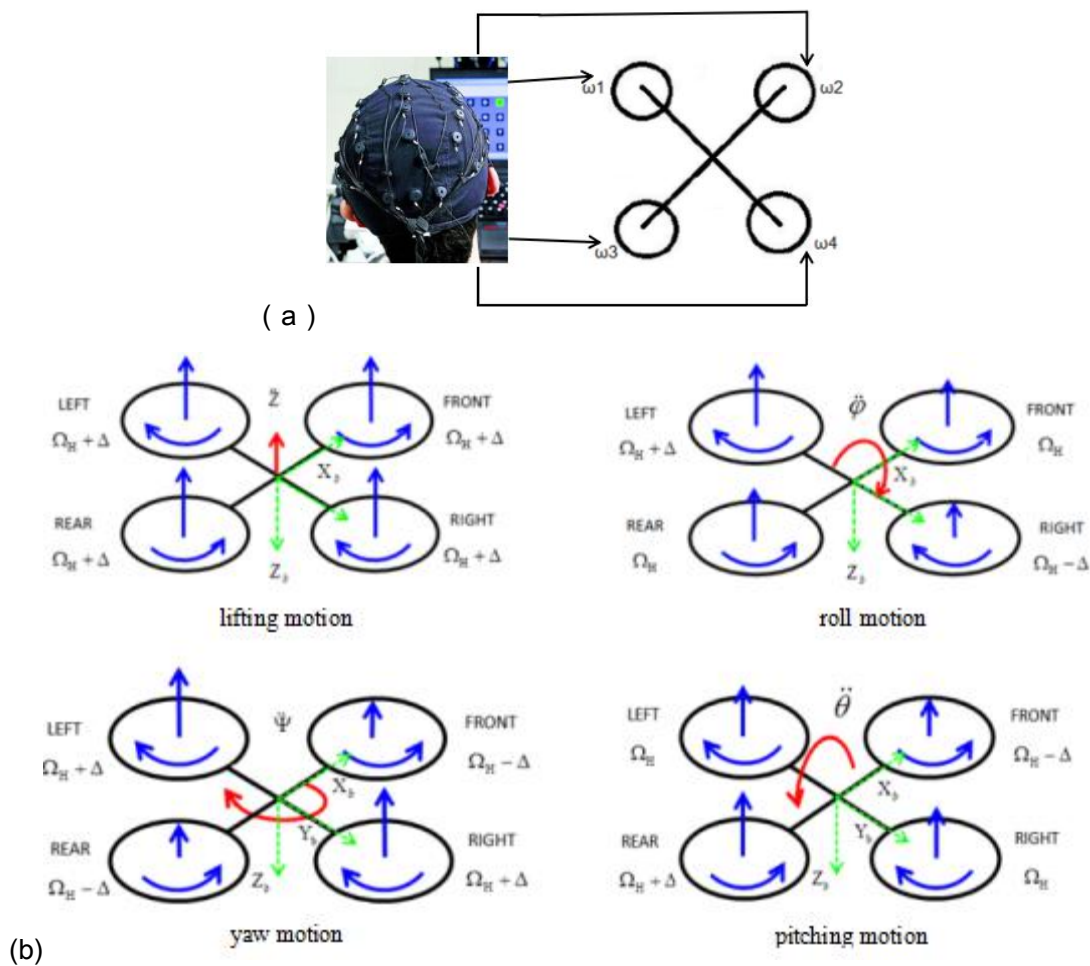

(a)

(b)

Fig 2 brain control combined with uav attitude control

## 4 auxiliary fuzzy PID controller

In this paper, fuzzy adaptive PID control algorithm is adopted to design a dual-loop control system to realize position control and attitude control of quadrotor. Control system structure is shown in figure. The outer ring adopts fuzzy PID control to control the space position of the quadrotor. The inner ring adopts the conventional PID to control the attitude of quadrotor, including yaw Angle, pitch Angle and roll Angle. The inverse solution module realizes the transformation from inertial frame to aircraft frame.

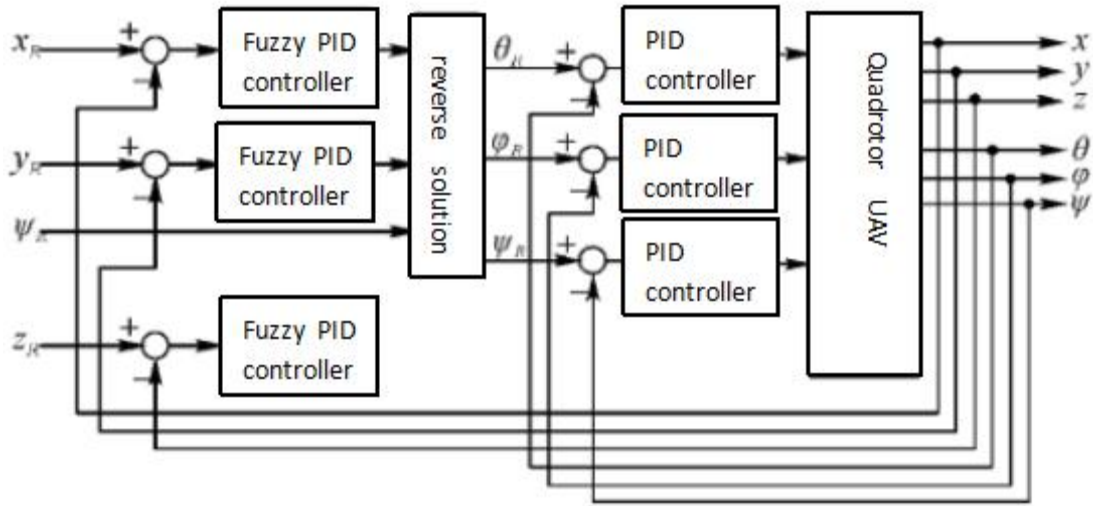

Fig 3 quadrotor UAV control system structure

Fuzzy PID controller uses fuzzy control rules to make fuzzy reasoning through current system error e and error change rate ec. The adjustment of parameters Kp, Ki and Kd was achieved by using the centroid method, and the whole adjustment process was divided into three stages.

(1) In the initial stage, if e is larger, take a larger Kp to improve the response speed. Set Ki = 0 to prevent integral saturation, and Kd is appropriate to reduce overshoot.

(2) In the middle stage of regulation, moderate Kp, Ki and Kd are adopted to ensure a certain response speed and avoid overshoot.

(3) In the later stage, reduce Kp appropriately to reduce the static difference, increase Ki appropriately to improve stability, and reduce Kd appropriately to prevent oscillation.

Kp, Ki and Kd all use triangular membership functions, and the fuzzy subset is {NB, NM, NS, ZO, PS, PM, PB}. Specific fuzzy control rules are shown in table 1, table 2 and table 3.

Table 1 Kp fuzzy rule table

| $e$ | ec | | | | | | |
|---|---|---|---|---|---|---|---|
| | NB | NM | NS | ZO | PS | PM | PB |
| NB | PB | PB | PB | PB | PB | PB | PM |
| NM | PM | PM | PM | PM | PM | PM | PM |
| NS | PS | PS | PS | PS | PS | PS | PS |
| ZO | PS | ZO | ZO | NS | ZO | ZO | PS |
| PS | PS | PS | PS | PS | PS | PS | PS |
| PM | PM | PM | PM | PM | PM | PM | PM |
| PB | PM | PM | PS | PS | PS | PM | PB |

Table 2 Ki fuzzy rule table

| $e$ | ec | | | | | | |
|---|---|---|---|---|---|---|---|
| | NB | NM | NS | ZO | PS | PM | PB |
| NB | NB | NB | NB | NB | NB | NB | NB |
| NM | NM | NM | NM | NM | NM | NM | NM |
| NS | NS | ZO | PM | PM | PM | ZO | NS |
| ZO | ZO | PS | PM | PB | PM | PS | ZO |
| PS | PM | ZO | PM | PM | PM | ZO | NS |
| PM | NM | NM | NM | NM | NM | NM | NM |
| PB | NB | NB | NB | NB | NB | NB | NB |

Table 3 Kd fuzzy rule table

| e | ec | | | | | | |
|---|----|----|----|----|----|----|----|
|   | NB | NM | NS | ZO | PS | PM | PB |
| NB | PB | PM | PS | PS | PS | PM | PM |
| NM | NS | NS | NS | NS | NS | NS | PM |
| NS | PS | NS | NM | PM | NM | NS | PS |
| ZO | PS | NS | NM | NM | NM | NS | PS |
| PS | PS | NS | NM | PM | NM | NS | PS |
| PM | PM | NS | NS | NS | NS | NS | NS |
| PB | PB | PM | PS | PS | PS | PM | PM |

# 5 simulation experiment

A rectangular trajectory tracking control experiment of the same height is designed for small quadrotor to verify the algorithm. The simulation experiment adopts the runway track of 5m in fixed height as the virtual experimental scene, as shown in figure 5. The trajectory is composed of two straight lines and two circular arcs. The distance from the starting point to the first arc is 200m straight, and the length of each arc is 157m. It is used to test the control characteristics of the straight line and curve of the brain-controlled uav.

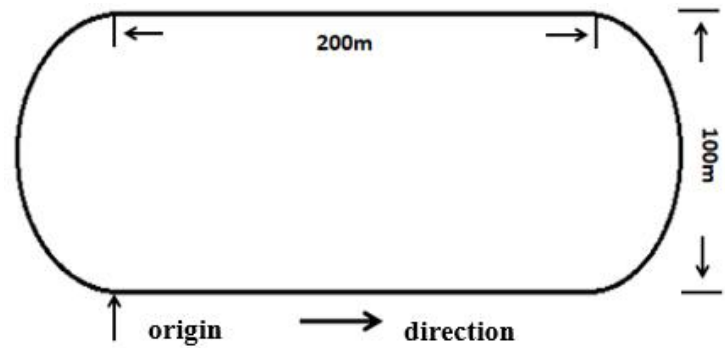

Figure 4 simulation experiment trajectory

参考文献

［1］ Nourmohammadi A, Jafari M, Zander T O. A Survey on Unmanned Aerial Vehicle Remote Control Using Brain–Computer Interface[J]. IEEE Transactions on Human-Machine Systems, 2018, 48(4):1-12.

［2］ Ming-tao Wang. Research on model predictive control methods for brain controlled vehicles [D]. Beijing institute of technology, 2016.

［3］ J. R. Wolpaw, N. Birbaumer, D. J. McFarland, G. Pfurtscheller, and T. M. Vaughan, "Brain–computer interfaces for communication and control," *Clinical Neurophysiol.*, vol. 113, no. 6, pp. 767–791, 2002.

［4］ H. Cecotti, "Spelling with non-invasive brain–computer interfaces— Current and future trends," *J. Physiol, Paris*, vol. 105, no. 1, pp. 106–114, 2011.

［5］ V. Gandhi, *Brain-Computer Interfacing for Assistive Robotics: Electroen- cephalograms, Recurrent Quantum Neural Networks, and User-Centric Graphical Interfaces*. New York, NY,

USA: Academic, 2014.

[6] Millán J R, Galán F, Vanhooydonck D, et al. Asynchronous non-invasive brain-actuated control of an intelligent wheelchair[C]. Engineering in Medicine and Biology Society, 2009. EMBC 2009. Annual International Conference of the IEEE. IEEE, 2009: 3361-3364.

[7] X. Gao, D. Xu, M. Cheng, and S. Gao, "A BCI-based environmental controller for the motion-disabled," [J]. IEEE Transactions on Neural Systems and Rehabilitation Engineering, vol.11,
no.2, pp.137–140, June 2003.

[8] D. J. McFarland and J. R. Wolpaw, "Brain-computer interface operation of robotic and prosthetic devices," Computer, vol. 41, no. 10, pp. 52–56, 2008.

[9] R. Prueckl, and C. Guger, "Controlling a Robot with a Brain-Computer Interface based on Steady State Visual Evoked Potentials," [C]. in 2010 Proc. International Joint Conference on Neural Networks (IJCNN), pp. 1-5.

[10] R. Ortner, C. Guger, R. Prueckl, E. Grünbacher, and G. Edlinger, "SSVEPs Based Brain-Computer Interface for Robot Control," [C]. in 2010 Proc. of the 12th international conference on Computers helping people with special needs, pp. 85-90.

[11] S. M. Torres Müller, T. F. Bastos-Filho, and M. Sarcinelli-Filho, "Using a SSVEPs-BCI to Command a Robotic Wheelchair," [C]. in Conf. Rec. 2011 IEEE International Symposium on Industrial Electronics (ISIE), pp. 957-962.

[12] M. Chung, W. Cheung, R. Scherer, and Rajesh P. N. Rao, "Towards Hierarchical BCIs for Robotic Control," [C]. in 2011 IEEE/EMBS 5th International Conference Neural Engineering Cancun, Mexico, pp. 330-333.

[13] L. Bi, X. Fan, N. Luo, K. Jie, Y. Li, and Y. Liu, "A head-up display-based P300 brain-computer
interface for destination selection," [J]. IEEE Trans. Intell. Transp. Syst., vol. 14, no. 4, pp. 1996–2001, 2013.

[14] Xin-an Fan. Research on human-computer interaction and control of brain-controlled vehicles [D]. Beijing: Beijing institute of technology, 2015

[15] D. Gohring, D. Latotzky, M. Wang, and R. Rojas, "Semi-autonomous car control using brain computer interfaces," [J]. in Advances in Intelligent Systems and Computing. Berlin, Germany: Springer-Verlag, 2013, pp. 393–408.

[16] A. S. Royer, A. J. Doud, M. L. Rose, and B. He, "EEG control of a virtual helicopter in 3-dimensional space using intelligent control strategies," *IEEE Trans. Neural Syst. Rehabil. Eng.*, vol. 18, no. 6, pp. 581–589, Dec. 2010.

[17] A. J. Doud, J. P. Lucas, M. T. Pisansky, and B. He, "Continuous three-dimensional control of a virtual helicopter using a motor im- agery based brain-computer interface," *PLoS One*, vol. 6, no. 10, 2011, Art. no. e26322.

[18] K. LaFleur, K. Cassady, A. Doud, K. Shades, E. Rogin, and B. He, "Quadcopter control in three-dimensional space using a noninvasive motor imagery-based brain–computer interface," *J. Neural Eng.*, vol. 10, no. 4, 2013, Art. no. 046003.